\documentclass[preprint,prb,aps,showpacs]{revtex4}

\newcommand{\bog}{B$_{1g}$}
\newcommand{\btg}{B$_{2g}$}
\usepackage{graphics}

\begin{document}

\title{Doping dependence of the superconducting gap in
  Bi$_2$Sr$_2$CaCu$_2$O$_{8 + \delta }$}
\author{K. C. Hewitt}
\altaffiliation[Work done while at]{Simon Fraser University,
Department of Physics} \affiliation{Department of Physics,
Dalhousie University, Halifax, N. S. Canada B3H 3J5}
\email{Kevin.Hewitt@Dal.ca}
\homepage{http://fizz.phys.dal.ca/~hewitt}

\author{J. C. Irwin} \affiliation{Department of Physics, Simon
Fraser University, Burnaby, B. C. Canada V5A 1S6}
\date{\today}

\begin{abstract}
Bi$_2$Sr$_2$CaCu$_2$O$_{8 + \delta }$ crystals with varying hole
concentrations (0.12 $< p < $ 0.23) were studied to investigate
the effects of doping on the {\it symmetry} and {\it magnitude} of
the superconducting gap ($\Delta ({\bf k})$). Electronic Raman
scattering experiments that sample regions of the Fermi surface
near the diagonal (B$_{2g}$) and principal axes (B$_{1g}$) of the
Brillouin Zone (BZ) have been utilized. The frequency dependence
of the Raman response function, $\chi^{\prime \prime }$, at low
energies ($\omega < \Delta({\bf k})$) is found to be linear for
B$_{2g}$ and cubic for B$_{1g}$ (T$<$ T$_c$).  The latter
observations have led us to conclude that the doping dependence of
the superconducting gap is consistent with d$_{x^2-y^2}$ symmetry,
for slightly underdoped and overdoped crystals. Studies of the
pair-breaking peak found in the B$_{1g}$ spectra demonstrate that
the magnitude of the maximum gap ($\mid\Delta_{max}\mid$)
decreases monotonically with increasing hole doping, for $p >
0.12$.  Based on the magnitude of the B$_{1g}$ renormalization, it
is found that the number of quasiparticles participating in
pairing increases monotonically with increased doping. On the
other hand, the B$_{2g}$ spectra show a weak "pair-breaking peak"
that follows a parabolic-like dependence on hole concentration,
for 0.12 $ < p <$  0.23.
\end{abstract}
\maketitle

\section{Introduction}

One of the important questions to be answered in identifying the microscopic
origin of superconductivity in cuprates is the nature of the
superconducting gap. A great deal of
attention \cite{ann,ding96,sta92,dev94,che,chen94,che94c,ope00,chen98,ken1}
has been focused on identifying
the symmetry of the pairing state since it is believed that this information
will provide constraints that will lead to the identification of the pairing
mechanism.  For example, it has been suggested \cite{mon91,mon92,mon92a,mon94,pine95,pin97}
that pairing in high temperature superconductors (HTSC) is induced by
the exchange of antiferromagnetic spin fluctuations and in this case the
pairing state would necessarily have d$_{x^2-y^2}$ symmetry.  From the
many experiments carried out in this context, a consensus view has emerged
\cite{sca95} that associates d$_{x^2-y^2}$ symmetry with the pairing state in
optimally hole-doped cuprates.  However, this consensus is lacking
(see Ref. \cite{timu98} for a review) in the results of Angle Resolved Photoemission
\cite{ding96,din95,loe96,shen97,din97,din98,mes99}, Tunneling \cite{miya98,ren98}, Infrared
\cite{puch96b} and Raman spectroscopy
\cite{ken1,blu97,blu97a,ken97,hew97,ein,hac96a,chen97a,dev96a,neme97,ope97,quil98,naei99,hew99}
experiments carried out on overdoped and underdoped compounds, where it appears that the
material properties are very different from those of the optimally doped
versions.  Since an appropriate model of the normal state must be capable
of predicting the effects of doping, it is clear that additional experiments
are required if one is to gain a complete understanding of superconductivity
in the high-T$_c$ cuprates.

Raman scattering has been widely used
\cite{sta92,dev94,che,chen94,che94c,ope00,chen98,ken1,hew97,ein,chen97a,irwin99,chen93}
to investigate electronic excitations in superconductors.  In electronic Raman
scattering experiments, one may investigate
various parts of the Fermi surface by a simple choice of the incident
and scattered polarization vectors.  The B$_{1g}$
spectra sample \cite{sta92,che,chen94,che94c}  regions of the Fermi surface located
near the Brillouin Zone
(BZ) principal axes [i.e. $(\pm \pi, 0),(0, \pm \pi )$], while B$_{2g}$ spectra sample regions
located near the diagonal directions in the BZ.  When Raman results are interpreted
\cite{sta92,dev94,che,chen94,che94c,ope00,chen98} in terms of the conventional model, \cite{kle84}
the gap is found to have maxima along the ($\pm \pi ,0$) and ($0, \pm \pi$)
directions and nodes (or minima) along the diagonal directions.  The results
are consistent with d$_{x^2-y^2}$ symmetry and are thus in
agreement with the outcome of several other experiments [see Scalapino \cite{sca95}
for a review].  The maximum values of the gap are found to scale with T$_c$
($2\Delta_{max} \sim 8k_BT_c$) between optimally doped
Bi$_2$Sr$_2$CaCu$_2$O$_{8 + \delta }$ (Bi2212) \cite{sta92},
YBa$_2$Cu$_3$O$_{7 - \delta }$ (Y123) \cite{che}, and La$_{2-x}$Sr$_x$CuO$_4$ (La214)
\cite{chen98}. In Bi2212, the value derived from Raman experiments ($ 2\Delta_{max} \sim 500
cm^{-1} $) is also in good agreement with the results of photoemission
experiments ($\Delta_{max} \approx 30 meV = 240 cm^{-1}$).

The first Raman investigation of the influence of doping was carried out
on Y123 \cite{che} and it was found that, although the symmetry appeared to be
independent of doping, the B$_{1g}$ pair-breaking frequency ($2\Delta_{max}$)
decreased much more rapidly than T$_c$ as the doping level
was increased beyond optimal.  Later, Kendziora {\it et al.} \cite{ken1}
carried out measurements on both heavily overdoped and underdoped Bi2212.  In
the heavily overdoped region they also found that $2\Delta /k_BT_c$ decreased
and that the B$_{1g}$ and A$_{1g}$ peak
frequencies approached equality, and suggested that this implied the
existence of an isotropic gap.  The low
energy frequency dependence of their measured B$_{1g}$ and A$_{1g}$ spectra also
appeared to be compatible with this interpretation.  However, it was pointed
out by Hewitt {\it et al.} \cite{hew97} that the gap cannot be isotropic if the
{\em Bose-factor corrected} spectra from all
scattering geometries are considered.  Hackl {\it et al.} \cite{hac96a}
also carried out experiments on a heavily overdoped crystal of Bi2212 and
interpreted their results in terms of a disordered d-wave model
proposed by Devereaux \cite{dev95}.

In an attempt to gain additional insight into the influence of doping on
the superconducting order parameter in HTSC we have carried out Raman scattering
experiments on overdoped (OD) and underdoped (UD) Bi2212 crystals with
critical temperatures, T$_{c}$ = 82K, 85K, 86K, 90.5K (UD) and 89K, 87K, 83K,
70K, 55K (OD).  Spectra have been obtained in the B$_{1g}$ and B$_{2g}$
scattering geometries in both the superconducting and normal
state.   The B$_{2g}$
spectra of overdoped compounds, which were not carefully investigated in
previous studies \cite{ken1}, have provided important information
regarding the symmetry of $\Delta ({\bf k})$.  These spectra sample
regions of the Fermi surface that
are located near the diagonal directions in the Brillouin zone and hence can
be used \cite{che,chen94,che94c} to directly probe for the existence of d-wave
type gap nodes.  Our results
suggest that the superconducting gap possesses d$_{x^2-y^2}$ symmetry,
for the doping range (0.12$ < p < $ 0.23) investigated -- slightly underdoped
to overdoped.

\section{Experimental Results}
\subsection{Sample Properties: Bi$_2$Sr$_2$CaCu$_2$O$_{8+\delta}$}
\label{PC-Bi2212}

In cuprates, the addition (removal) of oxygen increases
(decreases) the Cu valence \cite{mae:n=1-2-3} and therefore the
hole concentration.  Thus single crystals of Bi2212 were subjected
to various oxygen-rich or oxygen-poor environments in order to
vary the hole concentrations, spanning the doping range 0.07 $< p
< $ 0.23. The superconducting transition temperatures (measured by
the onset of the drop in magnetization or the zero of resistance)
and hole concentrations are listed in Table~\ref{bi2212samples}.
Hole concentrations for each crystal are estimated using the
empirical relation of Tallon and Presland \cite{tal95,pre91},
assuming T$_{c,max} = 92K$.

\subsection{Raman Scattering}
\label{R-Bi2212}

Raman vibrational spectra in the frequency range 20 - 1000 cm$^{-1}$
were obtained using the 514.5 nm line
of an argon ion (Ar$^{+}$) laser as the excitation source.  The Raman measurements
were carried out in a quasi-backscattering geometry, with the incident laser beam
directed nearly perpendicular to the freshly cleaved ({\bf a},{\bf b}) face of the crystal.

Bi2212 has an orthorhombic (D$_{2h}$) structure with the
{\bf a} and {\bf b} axes oriented at 45$^{\circ }$ to the Cu-O bonds.
In Raman experiments it is customary (since a$\approx$ b) to assume a tetragonal structure
(D$_{4h}$) with
axes parallel to {\bf a} and {\bf b}.
To facilitate comparison with other cuprates, however, we will consider
a tetragonal cell with {\bf x} and {\bf y} axes parallel to the
Cu-O bonds.  Another set of axes, {\bf x}' and {\bf y}' are rotated by
45$^{\circ }$ with respect to the Cu-O bonds. Considered within the
tetragonal point group (D$_{4h}$), the
{\bf x}'{\bf y}' (B$_{1g}$) scattering geometry means that the incident(scattered)
light is polarized along {\bf x}'({\bf y}') and selection of this
scattering channel enables coupling to excitations having B$_{1g}$ symmetry.
Similarly, the {\bf xy} geometry allows coupling to B$_{2g}$
excitations, which transforms as d$_{{\bf xy}}$.  Finally, the diagonal scattering
geometry {\bf x}'{\bf x}' allows coupling to A$_{1g}$ + B$_{2g}$ and
{\bf x}{\bf x} to A$_{1g}$ + B$_{1g}$ excitations.  Thus, by choosing
the polarization of the incident and scattered light appropriately, one
may select different components of the Raman tensor and thus various
symmetry properties of the excitations.

The B$_{1g}$ and B$_{2g}$ symmetries probe \cite{che,chen94,che94c} complementary
regions of the Fermi surface (FS),
while the A$_{1g}$ spectra correspond to an approximate average over the
entire FS.  In particular, the B$_{1g}$ Raman spectrum samples the FS
along the (0,$\pm \pi $) and ($\pm \pi $,0) directions whereas the
B$_{2g}$ spectum samples the ($\pm \pi$,$\pm \pi$)
directions.  That is, the B$_{1g}$ spectra sample the BZ principal
axes while the B$_{2g}$ spectra sample the BZ diagonals, and are
therefore used as  complementary probes to obtain information about
the {\bf k} dependence of the gap.

The spectra presented in this section have been corrected for the Bose
thermal factor.  Therefore, the spectra shown are proportional to the
imaginary part of the Raman response function ($\chi^{\prime \prime }$).

\subsubsection{B$_{1g}$ Renormalization}
\begin{figure}[ht!b!p!]
\includegraphics{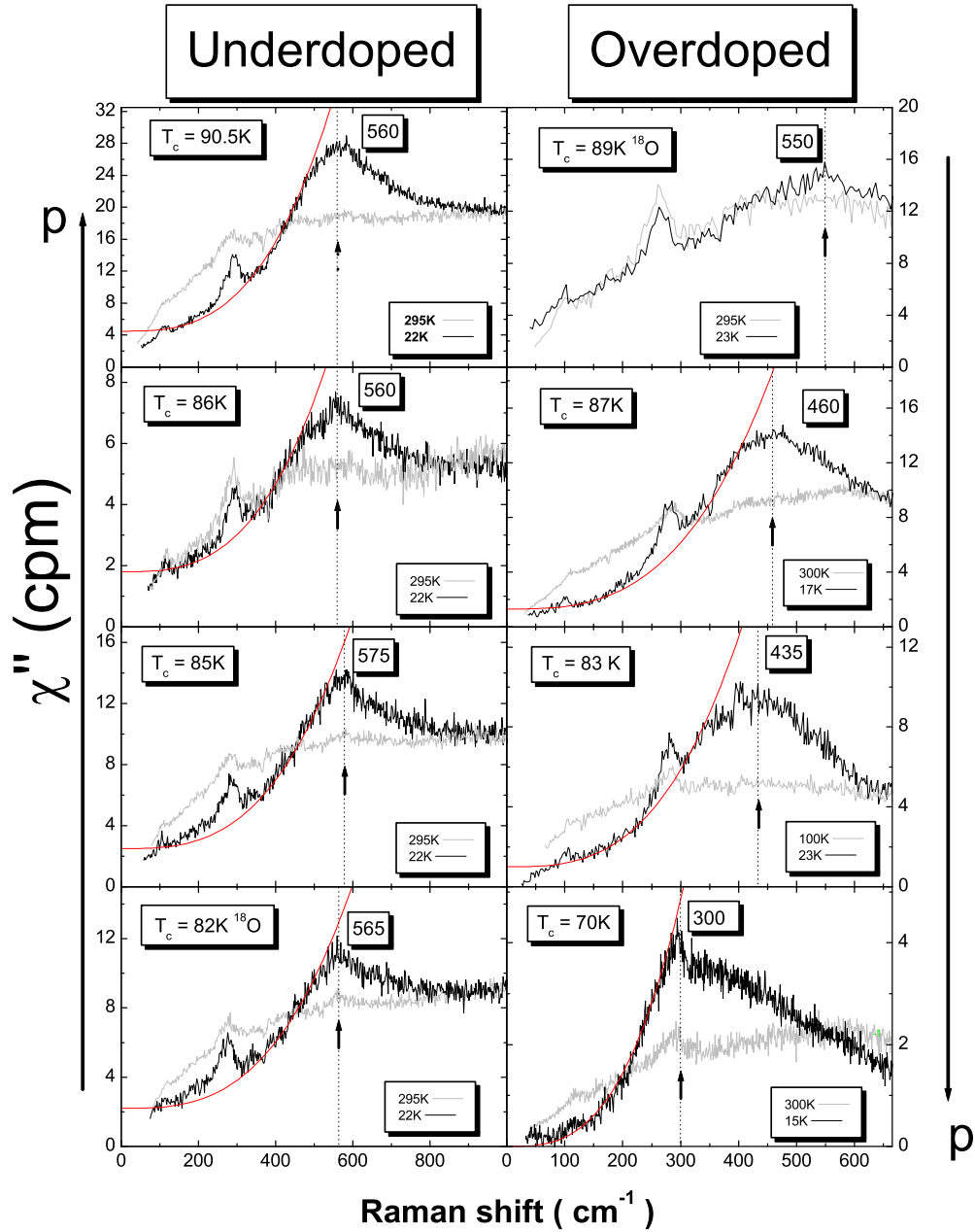}
\vspace{0.1in}
\caption{The $B_{1g}$ (${\bf x}'{\bf y}'$) spectra at
temperatures above and below T$_c$,
for underdoped samples with T$_c$'s of 82, 85, 86 and 90.5 K, and
for overdoped samples with T$_c$'s of 89, 87, 83 and 70 K.}
\label{b1g}
\end{figure}

\begin{figure}[ht!b!p!]
\includegraphics{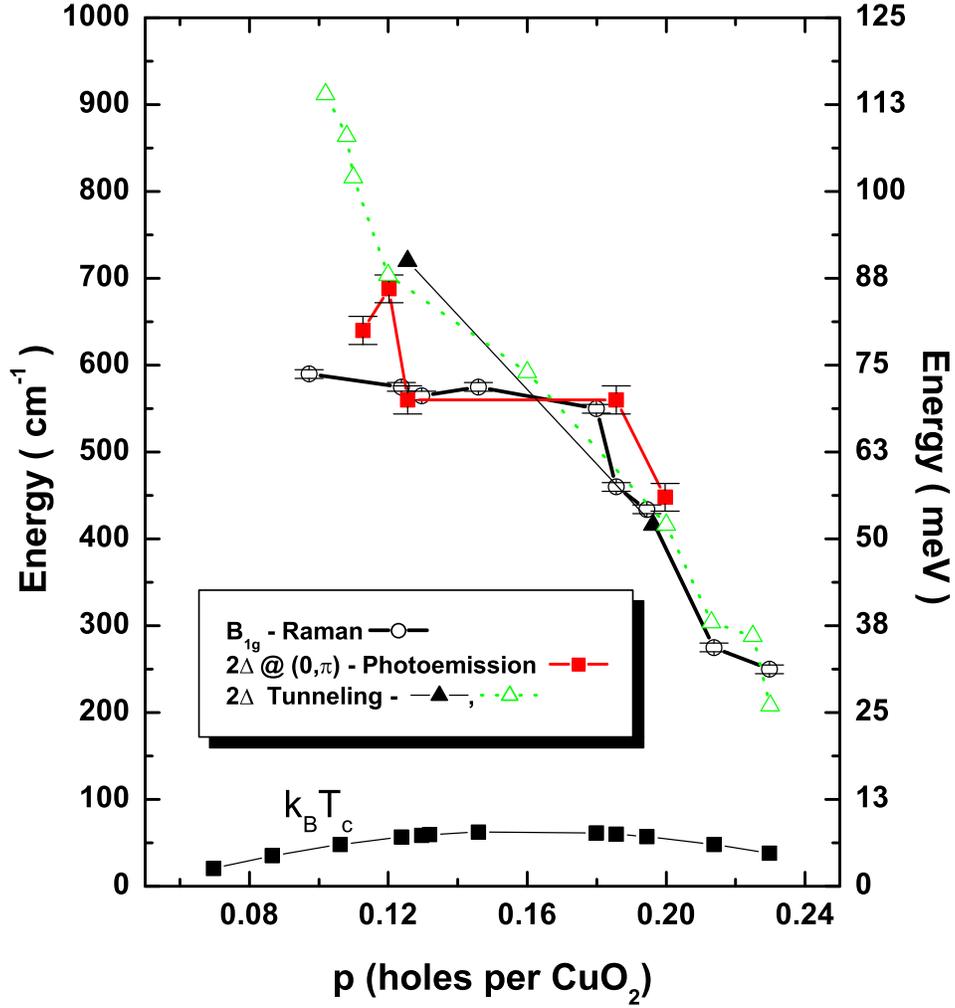}
\vspace{0.1in} \caption{The peak position of the superconductivity
induced redistribution in the B$_{1g}$ spectra as a function of
doping in Bi2212.  The corresponding values of $2\Delta_{max}$
derived from the ARPES results of Mesot {\it et al.} \cite{mes99}
(filled squares) , and the tunneling data of Miyakawa {\it et al.}
\cite{miya98} (filled triangles) and Ozyuzer {\it et al.}
\cite{ozy99} (open triangles) are shown for comparison. In all
cases Tallon's relation was used to determine the hole
concentration, assuming T$_{c,max} = 92K$.} \label{gap-B1g-ARPES}
\end{figure}

Spectra obtained from the cuprates at room temperature are
characterized \cite{sla91} by a rather featureless continuum, extending from $\hbar
\omega \sim$ 0 to 2 eV.  When the samples are cooled below T$_c$ the spectral
weight at low energies ($<  100 meV$ ) may be redistributed to higher
energy and gap-like peaks appear in the spectra.  In Bi2212 the superconductivity
induced renormalization observed in the
B$_{1g}$ spectra (see Fig.~\ref{b1g}) produces a peak whose energy remains
relatively constant (for p $<$ 0.16), but decreases drastically with increasing hole
concentration in the overdoped regime (p $>$ 0.16).  When the peak energy is
compared (Fig.~\ref{gap-B1g-ARPES})
with the maximum SC gap ($\Delta_{max}$) values reported by ARPES \cite{mes99} and
tunneling \cite{miya98,ozy99}, the results are striking.  The doping
dependence and energy of the
B$_{1g}$ peak position corresponds very well with the values of
$2\Delta_{max}$ obtained from ARPES measurements on similarly doped
samples (see Fig.~\ref{gap-B1g-ARPES}) .  This supports the previous
association of the peak  frequency in the B$_{1g}$ spectra with
the maximum ($2\Delta_o$) in a d$_{x^2-y^2}$ gap function. This
association is also supported by the low frequency dependence of
$\chi^{\prime \prime }$ at low temperatures, as will be discussed in section \ref{fdchi}.

\begin{figure}[ht!b!p!]
\includegraphics{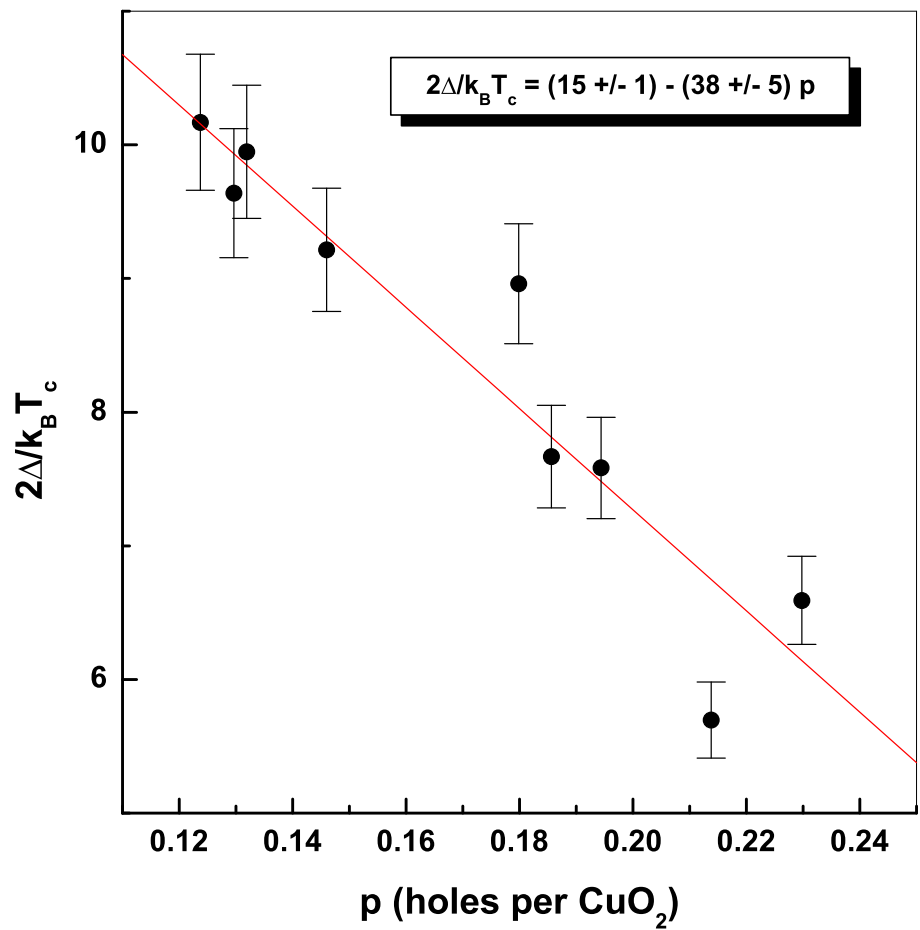}
\vspace{0.1in}
\caption{The hole concentration dependence of 2$\Delta_{max}/k_BT_c$.}
\label{kbtc-vs-p}
\end{figure}

The ratio $2\Delta_o/k_BT_c$ has also been plotted as a function of p in Fig.~\ref{kbtc-vs-p},
where the results reveal that this ratio decreases monotonically with
increasing hole concentration.  A linear least-squares fit to the data is well
described by the empirical relation,
\begin{eqnarray}
\frac{2\Delta_{max}}{k_BT_c} = & (15 \pm 1) - (38 \pm 5)p
\end{eqnarray}
for $0.12 < p < 0.24$.

\begin{figure}[ht!b!p!]
\includegraphics{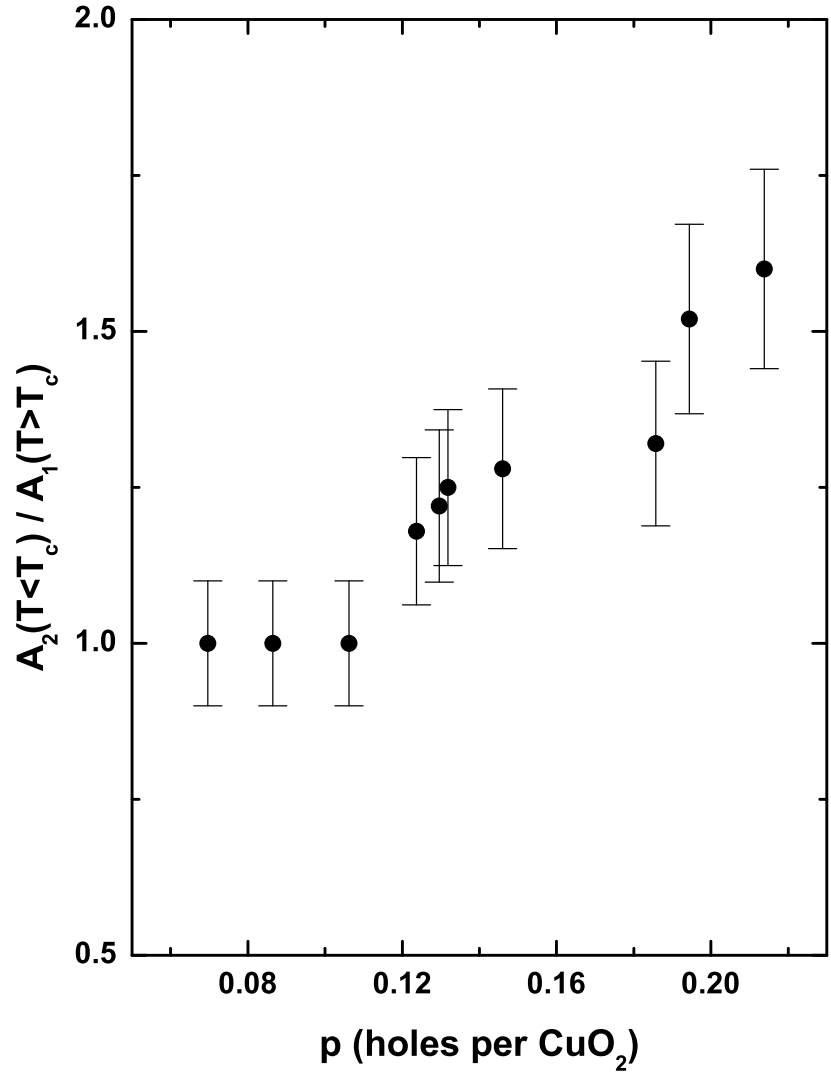}
\vspace{0.1in} \caption{The hole concentration dependence of the
ratio of the redistributed spectral weight in the normal and
superconducting state. The ratio is taken in the region of the
spectra where increased scattering is seen in the superconducting
state (see text). The lower three data points are taken from Ref.
\cite{hew99}.} \label{spectralweight}
\end{figure}

It can be seen in Fig.~\ref{b1g} that, as p is reduced, there is a
reduction in the fraction of spectral weight redistributed at low
frequencies. Quantitatively, this redistribution was estimated by
taking the ratio of the spectral weight above and below T$_c$. The
frequency range of interest is selected to envelope the region
where increased scattering in the superconducting state is
observed. Within this region, the points at which the normal and
superconducting state spectra cross defines the upper and lower
bounds on the integration. In this range, the ratio of the
integrated intensity of the superconducting (A$_2$) vs. normal
state (A$_1$) spectra are calculated, and the results are
presented in Fig.~\ref{spectralweight}.  The redistributed
spectral weight is reduced (Fig.~\ref{spectralweight}) with
decreasing hole concentration, indicating that fewer
quasiparticles are participating in superconductivity.  It may be
that the quasiparticle pairs are being destroyed by some
collective excitation of the crystal whose strength increases with
decreasing doping.  The missing spectral weight may be most likely
redistributed to higher energy, as recently suggested by the work
of Naeini {\it et al} \cite{nae00}. In their study of La214 it was
found that the B$_{1g}$ spectral weight that is lost at
low-energies ($\omega \le$ 800 cm$^{-1}$) is transferred to the
higher energy region normally occupied by multi-magnon scattering.
Taken together, the results suggest that the quasiparticle density
near $(\pi ,0)$ is reduced by magnetic excitations of some type.
Of course, this suggestion requires further quantitative analysis.

\subsubsection{B$_{2g}$ Renormalization}

\begin{figure}[ht!b!p!]
\includegraphics{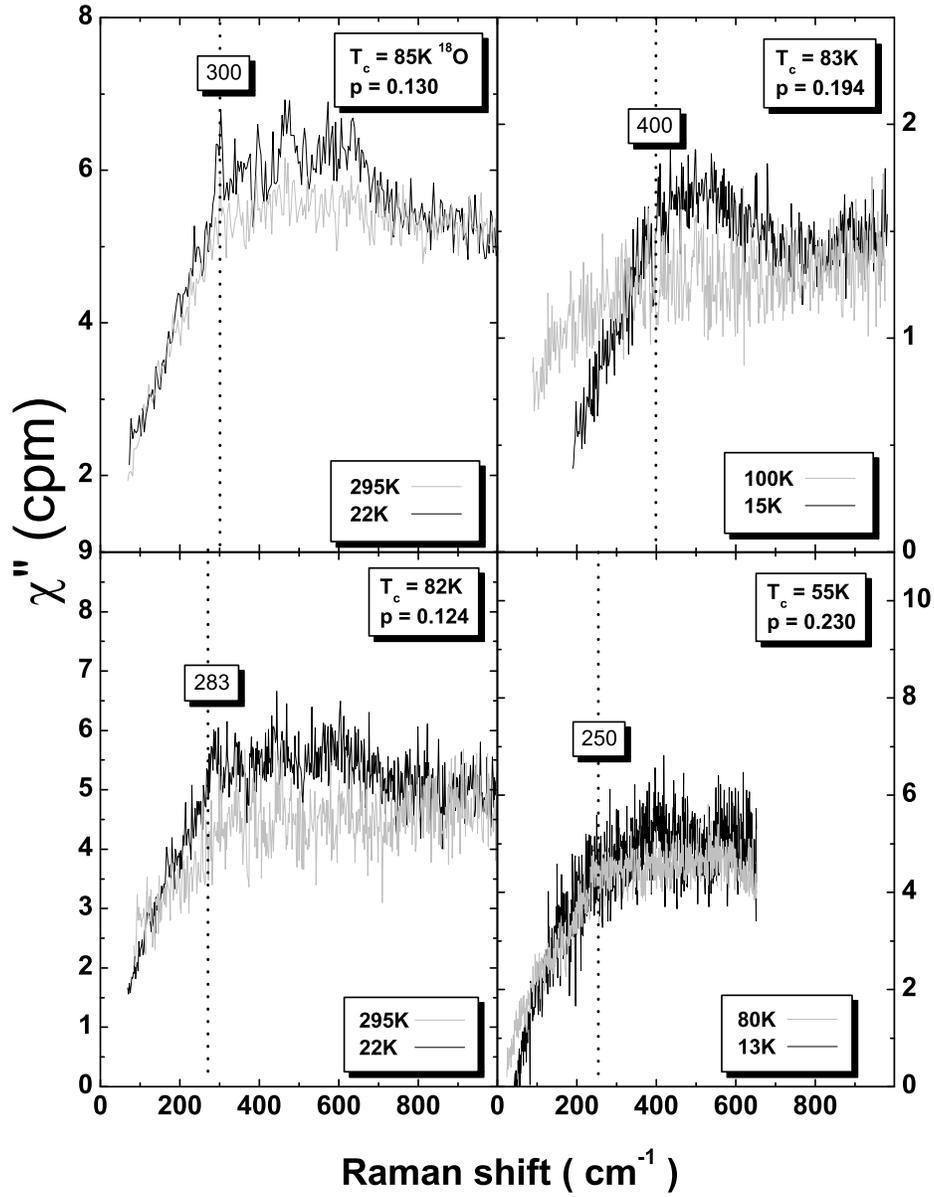}
\vspace{0.1in}
\caption{Doping dependence of the $B_{2g}$ ({\bf x}{\bf y}) Raman spectra at
temperatures above and below T$_c$,
for underdoped (UD) samples with T$_c$'s of 82 and 85 K, and for overdoped
(OD) samples with T$_c$'s of 83 and 55 K.}
\label{b2g}
\end{figure}

\begin{figure}[ht!b!p!]
\includegraphics{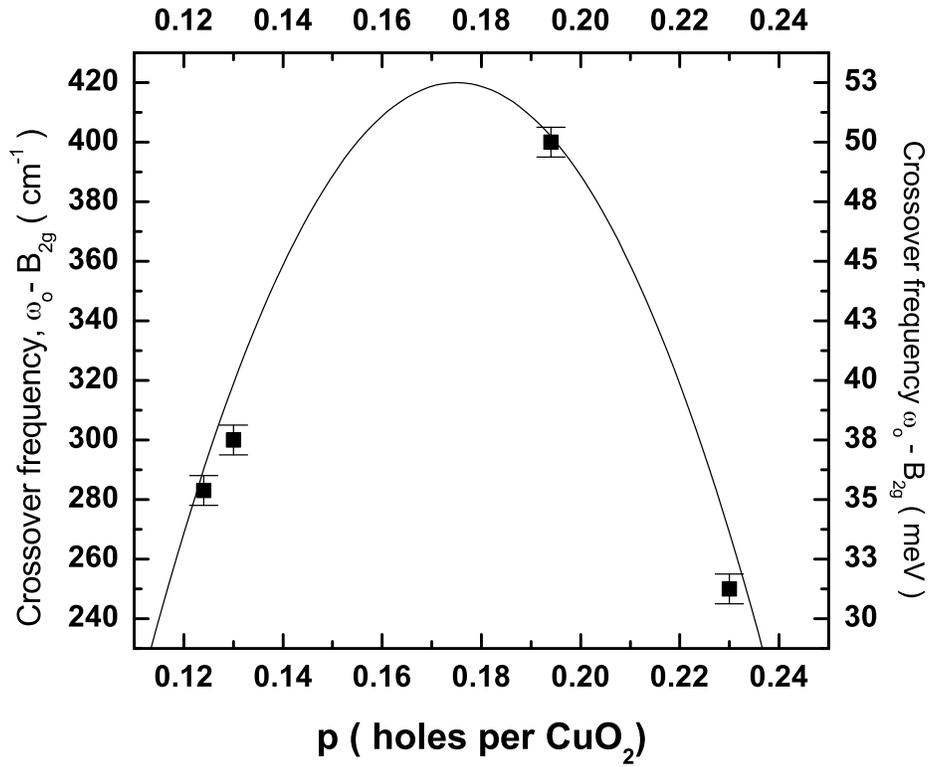}
\vspace{0.1in} \caption{The hole concentration dependence of the
linear-to-constant crossover position in the B$_{2g}$ spectra of
T$_c$ = 82, 85 K underdoped and T$_c$ = 83, 55K overdoped Bi2212
crystals.} \label{b2gpeak}
\end{figure}

A redistribution of spectral weight also occurs below T$_c$ in the
B$_{2g}$ channel (Fig.~\ref{b2g}) but it is difficult (except in
the case of the 83 K overdoped sample) to identify any peak
formation in the low temperature B$_{2g}$ spectra of
Fig.~\ref{b2g}. In the depleted regions of the spectra the
intensity rises linearly at low frequencies then levels off at a
frequency we will designate as $\omega_o$ (a linear-to-constant
crossover frequency). A linear fit is made to the low frequency
data and $\omega_o$ is defined by the point at which the line
departs from the experimental data points. This frequency can be
taken as a measure of the extent of the depleted region and thus
provides an estimate of the B$_{2g}$ "gap" energy \cite{chen94}.
It is interesting to note that $\omega_o$ has a parabolic
dependence on hole concentration (Fig.~\ref{b2gpeak}) and in
contrast to 2$\Delta_{max}$ (B$_{1g}$)actually scales with T$_c$
throughout the doping range, consistent with the results obtained
in Raman experiments by Opel {\it et al.} \cite{ope00} and Sugai
\& Hosokawa \cite{sug99}.  Thus, in the underdoped regime, the
results obtained seem to be incompatible with the existence of a
single, simple d-wave gap. On the other hand Sugai and Hosagawa
\cite{sug99} have suggested that the values of 2$\Delta_{max}$
obtained from the B$_{1g}$ spectra scale with the frequency of the
2-magnon peak. This observation suggests that B$_{1g}$ excitations
are magnetic in origin. Although not the topic of this paper, it
remains to be seen whether the holon-spinon phase diagram model of
Anderson \cite{and87,and87a,and91,and94} and Lee
\cite{nag90,nag92,wen96,lee97} may offer an explanation of these
results.

In the overdoped regime one can observe superconductivity induced
peaks in the B$_{1g}$ channel.  In this region, (p$\ge$ p$_{opt}$
both the values of $\omega_{peak}(B_{1g})$ and $\omega_o(B_{2g})$
decrease with increasing doping.  However, $\omega_{peak}(B_{1g})$
decreases much more rapidly than does $\omega_o(B_{2g})$ and as a
result the peak frequencies become almost equal for p $\approx$
0.25.  This behaviour, which is similar to that observed in doping
studies of La214\cite{naei99}, appears to be inconsistent
\cite{chen94} with the existence of a gap with d$_{x^2-y^2}$
symmetry.  Thus, although the B$_{1g}$ and B$_{2g}$ spectra of
optimally doped compounds \cite{chen94,dev94,chen98} imply the
existence of a d-wave gap, there are some puzzling discrepancies
in this interpretation in both the underdoped and overdoped
regions. In an attempt to gain additional insight we will now turn
to an investigation of the dependence of $\chi^{\prime \prime }$
on $\omega$ in the low energy region of the spectrum.

\subsubsection{Frequency Dependence of $\chi^{\prime \prime }$ at Low Energies.}
\label{fdchi}

\begin{figure}[ht!b!p!]
\includegraphics{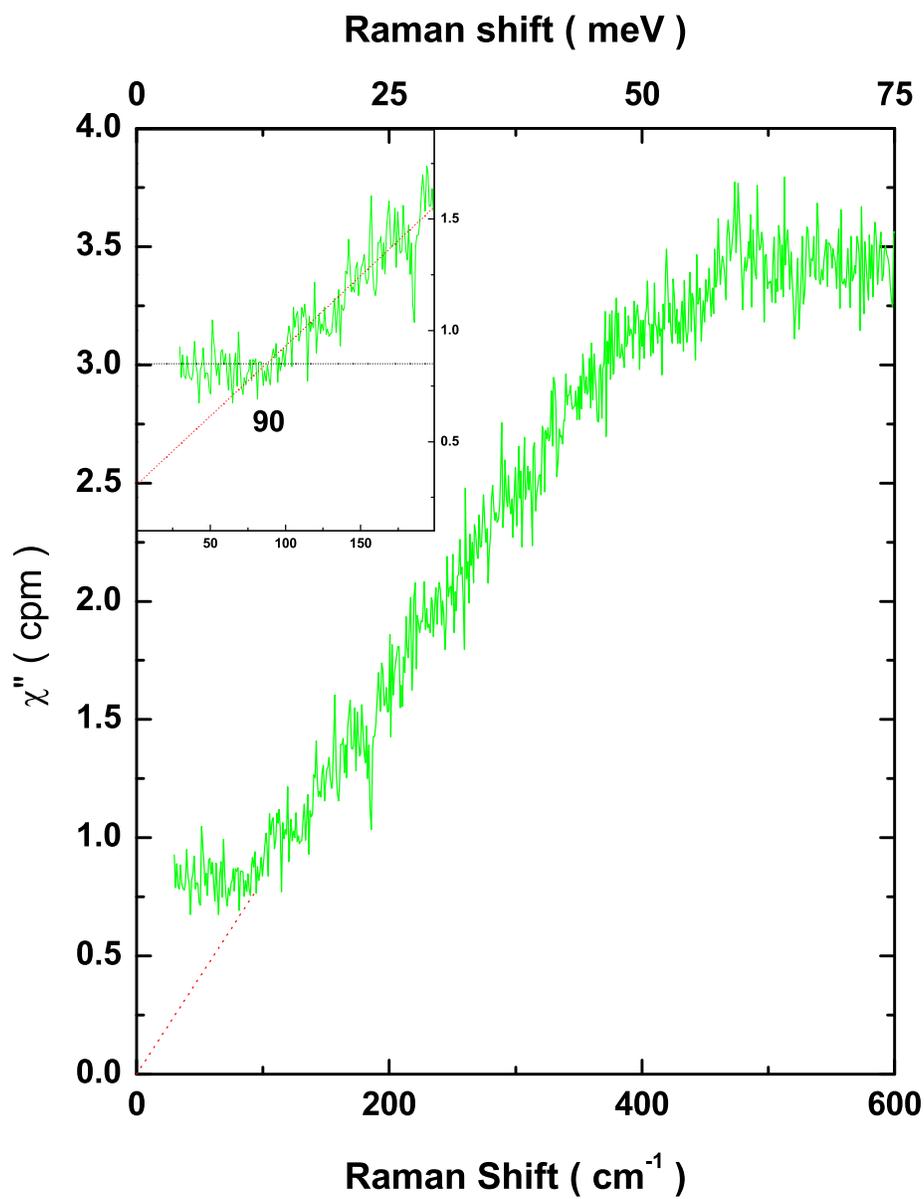}
\vspace{0.1in}
\caption{The B$_{2g}$ Raman spectra for a T$_c$ = 87K overdoped sample,
showing the low-frequency dependence in the inset.}
\label{87kb2g}
\end{figure}

The ($T/T_c \rightarrow 0$) frequency dependence of scattering in
the low energy regime,  (i.e. below $\omega_{peak}$) provides a
further test \cite{dev94,che} of the Cooper pair symmetry. In the
\btg\ geometry, the observed frequency dependence of the response
function remains linear in $\omega$, regardless of the sample
$T_{c}$ (Fig.~\ref{b2g}). The only exception is found in examining
the low energy region of the \btg\ spectra (Fig.~\ref{87kb2g}) for
the 87K (OD) sample. One observes a change in slope of the Raman
response function at an energy of $\omega \approx 90 cm^{-1} $ or
$\epsilon \approx 11 meV $. Below this energy the scattering is
constant and rises linearly at frequencies above 90 cm$^{-1}$.
These results have been repeated and are consistent with that
described here.  Thus we are confident that the observations are
intrinsic and not an experimental artifact. In an ARPES
investigation of a similarly overdoped (T$_C$= 87K) Bi2212
crystal, Ding {\it et al.} \cite{din95} found that the SC gap
function exhibited an extended node in the ($\pi ,\pi$), ($-\pi ,
-\pi$) directions in the BZ. The modulation direction corresponds
to the (0,0)$\rightarrow$ ($\pi , \pi$) region of the Fermi
surface of Bi2212. Ding {\it et al.} \cite{din96a} have proposed
that the extended node feature (p $\approx 0.186$) is an artifact
of umklapp scattering, resulting from scattering of the outgoing
electron by the wavevector of the modulation $q_M$ ($0.21\pi ,
0.21\pi$). The extended node artifact near ($\pi , \pi$) should be
manifest in the density of excited states by an increased
scattering at low energies.  Therefore, we contend that the
increased, and constant, scattering found at low energies in the
B$_{2g}$ channel (for this particular crystal) is consistent with
umklapp scattering from the modulation.  Umklapp processes scatter
a quasiparticle out of the first BZ. In this case, a quasiparticle
on the Fermi surface can be scattered by an integral multiple of
$q_M$.  Therefore, in the extended zone scheme, the quasiparticle
may thereby find itself on another portion of the Fermi surface
for which the Fermi velocity is opposite to that which it
possessed before being scattered.  Thus, this scattering process
may reverse the Fermi velocity of quasiparticles with wavevectors
along the direction of the modulation wavevector. This scattering
process may destroy the coherence that is required for pairing,
leading to a destruction or depression of the gap along the ($\pi
,\pi$) direction in k-space.  The effect would therefore be seen
in the B$_{2g}$ scattering geometry.  It is true that the Raman
intensity is proportional to the structure function $S(q,\omega
,T$.  If the scattering process is instantaneous, i.e. time
reversal symmetry holds, then the structure function is related by
the fluctuation-dissipation theorem to the susceptibility $\chi
\prime \prime$.  Time reversal symmetry implies that the $\chi (t)
= \chi (-t)$, or equivalently $\chi \prime \prime (\omega ) = \chi
\prime \prime (-\omega )$. Consequently, the value of $\chi \prime
\prime $ for $\omega \rightarrow 0$.........

In the \bog\ spectra of Fig.~\ref{b1g} the renormalization is
quite strong in the range of hole concentrations covering the
T$_c$'s 82 K (UD) to 70 K (OD) samples. In all but one spectra
(T$_c$ = 89 K OD) a cubic low frequency dependence is observed,
below the $2\Delta_o$ pair-breaking peak.

It is also evident that there is poorer agreement with a cubic fit
in the underdoped regime, where also the peak in the
superconducting state appears at a relatively constant value
around 565 $cm^{-1}$.  Recent tunneling spectroscopy studies of
underdoped Bi2212, which probe the gap at several points on
crystals show that there exists local variation in the gap size.
The results are interpreted in terms of an inhomogeneous
distribution of oxygen.  The poorer agreement (Fig.~\ref{b1g})
with a cubic fit is thought to be due to this intrinsic disorder,
giving rise to a linear component to the scattering at low
frequencies, as described by Devereaux et al \cite{dev95} and seen
in overdoped Bi2212. In addition, in other cuprate
\cite{neme97,ope00,chen97a,naei99,nae00,irwin99} supercondcutors
there is no redistribution of spectral weight below T$_c$ in the
underdoped regime, in contrast to the results presented here. This
contradiction is resolved by the assumption that the underdoped
crystals contains locally varying hole concentrations - the
fraction of the crystal with a local value of $p > 0.16$ is
reduced as the hole concentration is also reduced! Consequently,
the peak would remain near its value at optimal doping and be
reduced in intensity, as observed.

\subsubsection{Interpretation - Power Laws}

Pairing functions such as an anisotropic s-wave or a d-wave gap
are consistent with states lying below some
maximum gap value.  For a d-wave gap of the form d$_{x^{2}-y^{2}}$,
electronic Raman spectra in the B$_{2g}$ and B$_{1g}$ channels are predicted
\cite{dev94,che,chen94} to display a linear (in B$_{2g}$) and cubic
(in B$_{1g}$) power law behaviour below $\omega \sim \Delta$.
Conversely, the Raman response function in the low energy regime for an isotropic
s-wave gap should be \cite{dev95} characterized by the absence of
states below the gap, expressed in experimental terms by exponentially
activated behavior.

Down to the lowest frequency measured, the \bog\ and \btg\ spectra from all
samples show scattering below the gap peaks which appear in the spectra below
T$_c$. In particular, at all doping levels the B$_{2g}$ response
function shows a linear rise in scattering at low frequencies for T $<$ T$_c$.
Therefore, the results are consistent with any gap function that
posesses nodes (as a line or point) on the
Fermi surface near ($\pm \pi ,\pm \pi$).

In the hole concentration range $0.12 < p < 0.23$, the {\bf linear} and
{\bf cubic} frequency dependences found for the B$_{2g}$ (Fig.~\ref{b2g}) and
B$_{1g}$ (Fig.~\ref{b1g}) spectra, respectively, are consistent with
the behaviour predicted for a d$_{x^{2}-y^{2}}$ gap, and thus
completely inconsistent with an isotropic s-wave gap.

The cubic power law for a d$_{x^2-y^2}$ pairing state was derived in the
limit $\omega \rightarrow 0, T \rightarrow 0$. However, even at the
temperatures used in these studies, i.e 0.20 $<$ T/T$_c < $ 0.28, the
power laws are still consistent with a d-wave gap.  This suggests that the gap
opens quite quickly below T$_c$.

Despite the presence of cation disorder, oxygen intercalation, b-axis modulation,
and orthorhombicity that are common to Bi2212, the low-frequency power
laws appear to persist throughout the doping range. The low-frequency scattering
in the B$_{1g}$ channel is found to be consistent with a cubic frequency
dependence.  In B$_{2g}$ spectra, a linear-in-$\omega$ scattering is found
for all samples studied, which indicates that there are nodes in the gap
function near ($\pm \pi , \pm \pi$). Therefore, the combined (for $\omega < \omega_{peak}$)
power-law behavior ($\omega$ in B$_{2g}$ and $\omega^3$ in B$_{1g}$) points to the
existence of a d$_{x^2-y^2}$ gap function for the doping range studied.

\section{Discussion}
\label{scgd}

In the B$_{1g}$ scattering geometry, the superconductivity induced
renormalization peak frequency changes with doping.  Below a critical doping
level (p $\approx 0.12$) the renormalization is not present (see \cite{hew99}).
With increasing hole concentration  the peak energy remains relatively
constant to  p $\approx 0.18$ and
then drops rather abruptly, by approximately 50 \% over the hole concentration
range $0.18 \le p \le 0.23$ (i.e. overdoped regime). In the latter range,
the B$_{1g}$ peak energy and doping dependence can be plotted on a curve that
matches the $2\Delta_{max}$ feature observed in ARPES and tunneling spectroscopy
measurements.  Provided that the ARPES results do indeed measure
2$\Delta$, these considerations provide substantial support for the
assignment of the B$_{1g}$ renormalization peak to $2\Delta_{max}$.

\begin{figure}[ht!b!p!]
\includegraphics{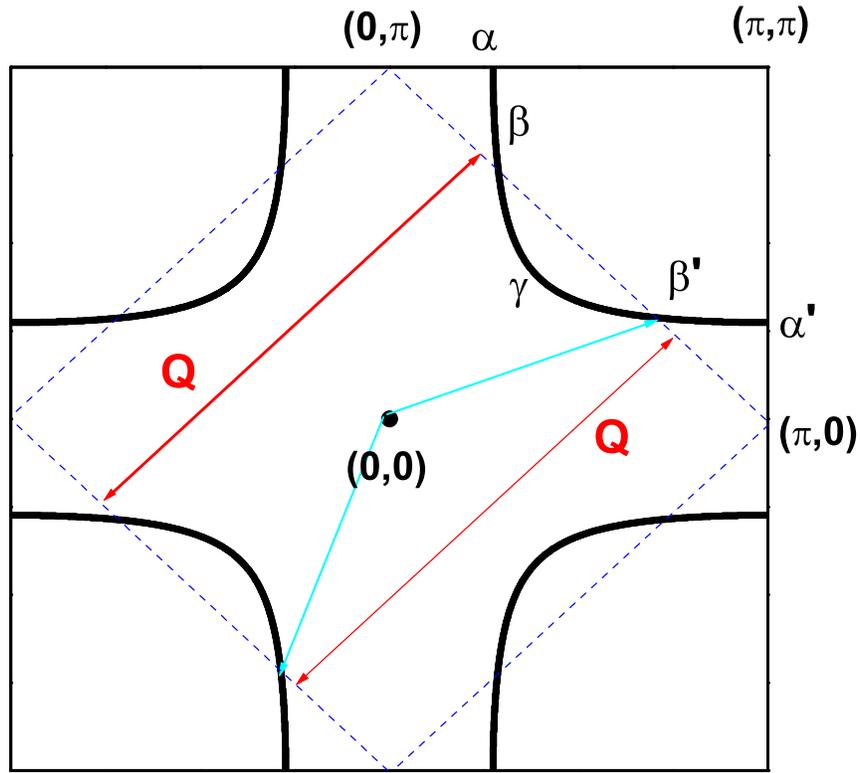}
\vspace{0.1in} \caption{A model of the
Fermi surface in cuprates (solid line) and the magnetic Brillouin
zone boundary (dashed line).  Quasiparticles near $\beta^{\prime
}$ {\it can} exchange an AFM wavevector ($\pi ,\pi$) while those
near $\gamma$ {\it cannot}.  Hence the former are highly scattered
"hot" quasiparticles, and the latter are referred to as "cold"
quasiparticles.} \label{hotspots}
\end{figure}

One reason for the lack of a pair-breaking peak relies on the fact that
the low-energy B$_{1g}$ spectral weight is reduced
\cite{chen97a,naei99,irwin99} substantially with underdoping.
If the quasiparticles are scattered away from the ($\pm \pi ,0$) regions
of the Fermi surface, then the resulting B$_{1g}$ spectra would
show a reduced intensity.  A mechanism for such anisotropic scattering lies
in the description of Raman scattering in the Nearly Antiferromagnetic Fermi
Liquid model \cite{pin97,dev97a}.  In the model, the presence of antiferromagnetic
spin fluctuations, which are strongly peaked near the
antiferromagnetic ordering vector $Q \approx (\pi , \pi)$, give rise to a
enhanced scattering of quasiparticles on regions of the Fermi
surface that are located near the BZ axes.  For example, in such a
process a quasiparticle can be scattered from ($\pi ,0$) to ($0,-\pi$)
by emitting a ($\pi ,\pi$) magnon and back to ($\pi ,0$) by emitting a
($-\pi ,-\pi$) magnon (Fig.~\ref{hotspots}).  Since the regions of the FS located near the
diagonals are far from resonance with Q, the quasiparticles in this
region are much more weakly scattered.  This gives rise \cite{pin97}
to the terms "hot-spots" for regions of the FS near the BZ axes and "cold-spots"
for those near the diagonals.  As the doping level decreases and the
antiferromagnetic fluctuations grow and become more strongly peaked
near Q, the scattering will increase and hence the B$_{1g}$ spectral
weight will decrease further.  Correspondingly, one would expect that
Cooper pairing of the highly damped "hot" quasiparticles to become
increasingly unlikely in the underdoped regime.  on this basis, the
absence of a pair-breaking peak in the B$_{1g}$ channel for low doping
levels (p$<$0.12) is thus to be expected.  This picture also accounts
for the rapid reduction, with decreasing hole concentration, of the
B$_{1g}$ spectral weight that is involved in superconductivity.

It is also interesting to note that the B$_{1g}$ spectra obtained from
underdoped Bi2212 appear to differ from that obtained from underdoped
La214 and Y123.  In the latter two compounds the B$_{1g}$ spectra
change much more abruptly when the doping level is reduced below
optimum.  In these compounds a strong 2$\Delta_{max}$ peak is observed
in the B$_{1g}$ spectra obtained from optimally doped materials, but
such a peak is completely absent
\cite{chen93,neme97,chen97a,naei99,irwin99} in the underdoped
compounds.  In Bi2212 however, a relatively strong B$_{1g}$ peak is
observed still observed in the underdoped compounds (see \cite{sug99}
and Fig.~\ref{gap-B1g-ARPES}).  In addition, the B$_{1g}$ spectral
weight drops rapidly as one enters the underdoped regime
\cite{chen97a,naei99,irwin99} of La214 or Y123 but decreases more
slowly with underdoping in Bi2212. It thus appears that the optimally
doped state is more sharply defined in both La214 and Y123 than it is
in Bi2212.  This might be attributed to the possibility that, because
of cation disorder and the superstructural modulation, Bi2212 is more
inhomogeneous than La214.

\section{Conclusions}

The hole concentration in single crystals of Bi2212 has been varied
from 0.10 to 0.22 by varying the oxygen concentration and by Y
substitution.  It has been found that the 2$\Delta_{max}$ peak in the
B$_{1g}$ spectrum increases in energy as the hole concentration is
reduced from p=0.22 to p=0.12 according to the empirical relation,
$2\Delta /k_BT_c = (15 \pm 1) - (38 \pm 5)p$.  Both the magnitude of
$\Delta_{max}$ and its' doping dependence, as determined from the
Raman spectra, are in good agreement with the results of ARPES
measurements \cite{mes99} of the leading edge shift that occurs below
T$_c$.  These results provide support for the association of the
2$\Delta$ peak of the B$_{1g}$ spectra with the superconducting gap
energy as determined by ARPES measurements .  It is also found that
the strength of the B$_{1g}$
renormalization, that occurs below T$_c$, becomes weaker as the hole
concentration is reduced from p=0.22 to p=0.12 and cannot be observed
in crystals with hole concentrations less than p$\approx$0.12.  This
observation is consistent with the selective scattering of
quasiparticles on regions of the Fermi surface that are located near
the k$_x$ and k$_y$ axes of the Brillouin zone.  A possible mechanism
for this scattering is provided by antiferromagnetic spin fluctuations
\cite{pin97,dev97a,naei99,irwin99} which grow in strength as the hole
concentration of the crystals is decreased.

The B$_{2g}$ spectra show a very weak renormalization at T$_c$ for
two of the four crystals studied in this work. The 82K The
cross-over frequency ($\omega_o$) exhibits a parabolic dependence
on hole concentration - similar to that found for T$_c$. In fact
$\omega_B$ appears to scale with T$_c$ according to the relation
$\hbar \omega_B$ $\approx$ 5k$_B$T$_c$, as has been suggested by
others \cite{neme97,chen97a,irwin99}. The B$_{2g}$ response
function, at T = 20K, increases linearly at low frequencies in
most of the samples studied.   This observation is consistent
\cite{dev94} with the presence of nodes in the superconducting
gap, in the direction $|k_x|=|k_y|$.  The only exception to this
linear behaviour at low temperature was observed in the B$_{2g}$
spectrum of an overdoped crystal with T$_c$ =87K. In this case the
scattered intensity remained constant for all frequencies $\omega
\le 90$ cm$^{-1}$.  It is interesting that in ARPES experiments
carried out on an overdoped crystal with the same T$_C$ it was
found that gap was zero over an extended region of the FS, which
could be consistent with $\chi^{\prime \prime } \approx $ constant
below a certain frequency.  This constant is believed to be an
artifact associated with umklapp scattering from the
superstructural modulation.

Finally, it should be noted that results obtained here for the doping
dependence of the spectra are, in most respects, similar to those
obtained in other cuprates \cite{neme97,ope00,chen97a,naei99,nae00,irwin99}.
One exception appears to be the fact that in the underdoped regime the spectral
weight depletion, and the strength of the B$_{1g}$ renormalization in Bi2212,
decrease less abruptly as p is decreased below p$\approx$
p$_{opt}$.

\section{Acknowledgements}
The financial support of the Natural Sciences and Engineering Research
Council of Canada is gratefully acknowledged. We are indebted to
D. M. Pooke of Industrial Research Ltd. (Lower Hutt, New Zealand) for
some of the single crystals used in these studies.

\bibliographystyle{prsty}

\begin{table}[h!!tb]
\caption{Bi2212 sample properties - see Ref. \cite{hew00} for
details.}
\begin{tabular}{|l|c|c|c|}
\hline Sample          & T$_c$  & Doping  & p        \\ \hline
\hline Bi-Y(0.40)-2212 &  30K   &  UD     & 0.0696   \\ Bi2212 &
51K   &  UD     & 0.0865 \\ Bi2212-$^{18}$O &  51K   &  UD     &
0.0865 \\ Bi-Y(0.07)-2212 &  70K   &  UD     & 0.1062 \\
Bi2212-$^{18}$O & 82K   &  UD     & 0.1237 \\ Bi2212          &
85K   &  UD     & 0.1296 \\ Bi2212          &  86K   &  UD     &
0.1319 \\ Bi2212 & 90.5K &  UD     & 0.1460 \\ Bi2212-$^{18}$O &
89K   &  OD & 0.1799 \\ Bi2212          &  87K   &  OD     &
0.1857 \\ Bi2212 & 83K   &  OD     & 0.1944 \\ Bi2212          &
70K   &  OD & 0.2138 \\ Bi2212          &  55K   &  OD     &
0.2298 \\ \hline
\end{tabular}
\label{bi2212samples}
\end{table}
\begin{table}[h!!tb]
  \centering
  \caption{Values of the cubic (b) and constant (a) terms used in a fit ($I = a + b\omega^3$) to the low frequency spectra of
  Fig.\ref{b1g} - UD = Underdoped; OD = Overdoped}
\begin{tabular}{|c|c|c|c|c|} \hline
  p      &  OD or UD & T$_c$ (K) & a    & b \\ \hline \hline
  0.124  &  UD       & 82        & 2.2  & $0.06 X 10^{-6}$    \\
  0.130  &  UD       & 85        & 2.5  & $0.07 X 10^{-6}$    \\
  0.132  &  UD       & 86        & 1.8  & $0.05 X 10^{-6}$    \\
  0.146  &  UD       & 90.5      & 4.5  & $0.18 X 10^{-6}$    \\
  0.186  &  OD       & 87        & 1.3  & $0.18 X 10^{-6}$    \\
  0.194  &  OD       & 83        & 1.0  & $0.18 X 10^{-6}$    \\
  0.214  &  OD       & 70        & 0.0  & $0.18 X 10^{-6}$    \\
  \hline
\end{tabular}

  \label{fit-co-efficients}
\end{table}

\end{document}